\documentclass[aps,manuscript,showpacs,showkeys,superscriptaddress]{revtex4-1}
\usepackage{graphicx}% Include figure files
\usepackage{epsfig}  % Include figure files
\usepackage{epsf}    % Include figure files
\usepackage{dcolumn}% Align table columns on decimal point
\usepackage{bm}% bold math
\usepackage{dcolumn}% Align table columns on decimal point
\usepackage{textcomp}% Align table columns on decimal point
\usepackage[tbtags]{amsmath}
\usepackage{amsfonts}
\usepackage{float}
\usepackage{subfig}
\usepackage[]{hyperref}
  \hypersetup{
  unicode=false,          % non-Latin characters in Acrobat?s bookmarks
  pdftoolbar=true,        % show Acrobat?s toolbar?
  pdfmenubar=true,        % show Acrobat?s menu?
  pdffitwindow=true,     % window fit to page when opened
  pdfstartview={FitH},    % fits the width of the page to the window
  pdfsubject={Four photon absorption},   % subject of the document
  pdfnewwindow=true,      % links in new window
  pdfcreator={RevTeX},
  colorlinks=true,       % false: boxed links; true: colored links
  linkcolor=red,          % color of internal links
  citecolor=blue,        % color of links to bibliography
  urlcolor=blue,           % color of external links
  }
\usepackage{hypcap}

\begin{document}
% \ifpdf
%\DeclareGraphicsExtensions{.pdf,.jpg,.mps,.png}
% \else
%\DeclareGraphicsExtensions{.eps,.ps}
% \fi

%\preprint{TIFR/TH/12-03}

\title{Observation of four-photon absorption and determination of corresponding nonlinearities in CdS quantum dots}


%%%%%%%%%%%%%%%%%%%%%%%%%%%%%%%%%%%%%%%%%%%%%%%%%%%%%
\author{P. Ghosh \footnote{Corresponding author}}
\email[Email Address: ]{ghosh.pintu@iitb.ac.in}
\affiliation{Department of Physics, Indian Institute of Technology Bombay, Mumbai 400076, India}

\author{E. Ramya}
\affiliation{School of Physics, University of Hyderabad, Hyderabad, India}

\author{P. K. Mohapatra}
\affiliation{Department of Physics, Indian Institute of Technology Bombay, Mumbai 400076, India}

\author{D. Kushavah}
\affiliation{Department of Physics, Indian Institute of Technology Bombay, Mumbai 400076, India}

\author{D. N. Rao}
\affiliation{School of Physics, University of Hyderabad, Hyderabad, India}

\author{P. Vasa}
\affiliation{Department of Physics, Indian Institute of Technology Bombay, Mumbai 400076, India}

\author{K. C. Rustagi}
\affiliation{Department of Physics, Indian Institute of Technology Bombay, Mumbai 400076, India}

\author{B. P. Singh}
\affiliation{Department of Physics, Indian Institute of Technology Bombay, Mumbai 400076, India}
%\collaboration can be followed by \email, \homepage, 
%\thanks as well.
\date{\today}
%%%%%%%%%%%%%%%%%%%% abstract %%%%%%%%%%%%%%%%%%%%%%%%%%%%%%%%%%%%%%%%%%
\begin{abstract}
Bound- and excited-state electronic nonlinearities in CdS quantum dots 
have been investigated by Degenerate Four-Wave Mixing (DFWM) and Z-scan techniques in 
the femtosecond time regime. 
This QD sample shows Kerr-type nonlinearity for incident beam intensity 
below 0.18 TW/cm$^2$. However, further increment in intensity results in four-photon absorption (4PA) 
indicated by open- and closed-aperture Z-scan experiments. 
Comparing open-aperture Z-scan experimental results with theoretical models, the 4PA coefficient 
$\alpha_4$ has been deduced. Furthermore, third-order nonlinear index $\gamma$ and 
refractive-index change coefficient $\sigma_r$ corresponding to excited-state electrons due to 4PA 
have been calculated from the closed-aperture Z-scan results. 
UV-visible absorption and photoluminescence 
experimental results are analyzed towards estimating band gap energy and defect state energy. 
Time Correlated Single Photon Counting (TCSPC) was employed to determine the decay 
time corresponding to band-edge and defect states.
The linear and nonlinear optical techniques have allowed the direct observation of 
lower and higher-order electronic states in CdS quantum dots.
\end{abstract}
%%%%%%%%%%%%%%%%%%%%%%%%%%%%%%%%%%%%%%%%%%%%%%%%%%%%%%%%%%%%%%%%%%%%%%%%%%
%\maketitle must follow title, authors, abstract, 
%\pacs, and \keywords
\maketitle

%%%%%%%%%%%%%%%%%%%%%%%%%%  body  %%%%%%%%%%%%%%%%%%%%%%%%%%
\section{Introduction}
The physics of quantum dots (QDs) is of great scientific interest from both fundamental 
and application point of view.
A comprehensive knowledge about nonlinear absorption and refraction processes in 
quasi-zero dimensional semiconductor structures or QDs  
is important for further development of
nonlinear-optical semiconductor devices \cite{Marcelo2013, Dakovski2013, Lad2007, Guang2008}. 
Quest for knowledge about this topic can be adequately addressed by nonlinear optical 
experimental techniques, such as Z-scan \cite{YoshinoPhysRevLett.91.063902, Sheik-Bahae1990, Said1992, Wei1992}, 
degenerate four-wave mixing (DFWM) \cite{Canto-Said1991, Bindra1999}, 
and pump-probe spectroscopy \cite{Gaponenko1994}. 
Over the past years, these nonlinear optical 
experimental techniques have been extensively used as powerful tools towards investigating 
the excited electron-hole pair states dynamics of semiconductor QDs, providing complementary 
information that obtained by linear optical experimental techniques. 
With the access of ultrafast and ultrahigh intense laser pulses, multiphoton 
absorption $\it i.e.$ simultaneous absorption of two or more photons has been 
extensively studied. 
These multiphoton absorption processes are exceedingly promising 
in many fields including
optical limiting \cite{He:95, Prasad2008, Venkatram2008, Kiran:02}, 3D microfabrication \cite{Maruo:97}, 
optical data storage \cite{Nature2002PNPrasad, PARTHENOPOULOS1989}, and
biomedical applications \cite{Yanik2006}. In this regard, CdS QDs are of particular interest because 
of their high intrinsic nonlinearity \cite{Kalyaniwalla1990}.\par
So far, various nonlinear processes for comprehensive materials 
were studied \cite{Sheik-Bahae1990, Canto-Said1991, Said1992}. 
Furthermore, 
third-order nonlinear index $\gamma$ and 
refractive-index change coefficient $\sigma_r$ corresponding to free-carriers due to TPA 
have been calculated from closed-aperture Z-scan results \cite{Said1992}. 
To the best of our knowledge, there are hardly any work included discussion on deriving these nonlinear 
parameters for three or four-photon absorption in QDs.\par
In this paper, we report the detail investigation of nonlinear optical processes in 
CdS QDs synthesized by gamma-irradiation technique. Towards understanding these processes, 
intensity dependent DFWM, open, and closed-aperture Z-scan experiments were performed. 
Furthermore, we derived $\gamma$ and $\sigma_r$ values corresponding to excited-state electrons generated 
by four-photon absorption. 
Results of open-aperture Z-scan with 400 nm femtosecond laser pulses has also been presented.
In the first section of results and discussion, we report nonlinear studies on this CdS QD sample. 
In the later part, we present UV-visible absorption, room temperature 
photoluminescence and TCSPC experimental results for better understanding of the electronic states 
in the QDs. 
\section{Experimental}
The results of ultrafast nonlinear experiments including DFWM, 
open-aperture and closed-aperture Z-scan on colloidal 
solution of CdS QD sample have been reported in this paper. 
These nonlinear studies are performed using a Ti: Sapphire femtosecond laser
(Spectra-Physics, Mai Tai, Spitfire amplifier) having wavelength $\lambda = 800$ nm, 
and repetition rate 1 KHz. 
The pulse width was determined to be 110 fs through intensity autocorrelation measurements. 
The nonlinear properties are investigated for the intensity 
regime 0.02 TW/cm$^2$ to 0.80 TW/cm$^2$ with the femtosecond laser pulses. 
The input beam intensity 
is varied using a polarizer and a $\lambda/2$ plate combination.
It can be noted that at this intensity range, the 
water solution does not show any nonlinear behaviour for DFWM as well as Z-scan experiments.
The DFWM experiments are performed using folded boxcar geometry \cite{Wise1998}. In this technique, 
a three-dimensional phase-matching is implemented, which enables spatial separation of 
the signal-beam from the input beams. 
The fundamental beam is divided into three nearly equal
intensity beams (intensity ratio of 1:1:0.9) in such a way that they form
three corners of a square and are focused into the nonlinear medium. 
All three beams are synchronized both spatially and temporally. 
The resultant DFWM signal is generated due to the phase-matched interaction: 
$\overrightarrow{k}_4=\overrightarrow{k}_1-\overrightarrow{k}_2+\overrightarrow{k}_3$. 
In Z-scan experiments, a Gaussian 
laser beam is tightly focused onto an optically non-linear sample
using a finite aperture and the transmittance through 
the medium is measured in the far field. 
Finally, the resultant transmittance 
is recorded as function of the sample position Z measured about the focal plane.
Open-aperture Z-scan has also been performed at wavelength 400 nm (second harmonic of 
the fundamental wavelength from a BBO crystal). 
The details about synthesis and structural characterization of 
the CdS QDs are reported in \cite{Soumyendu2012}.
Particle size distribution and chemical composition are obtained from 
the HRTEM images, XPS and Raman spectra analysis.
\section{Results and discussion}
\begin{figure}[H]
\centering
  \includegraphics[height=0.25\textheight,keepaspectratio]{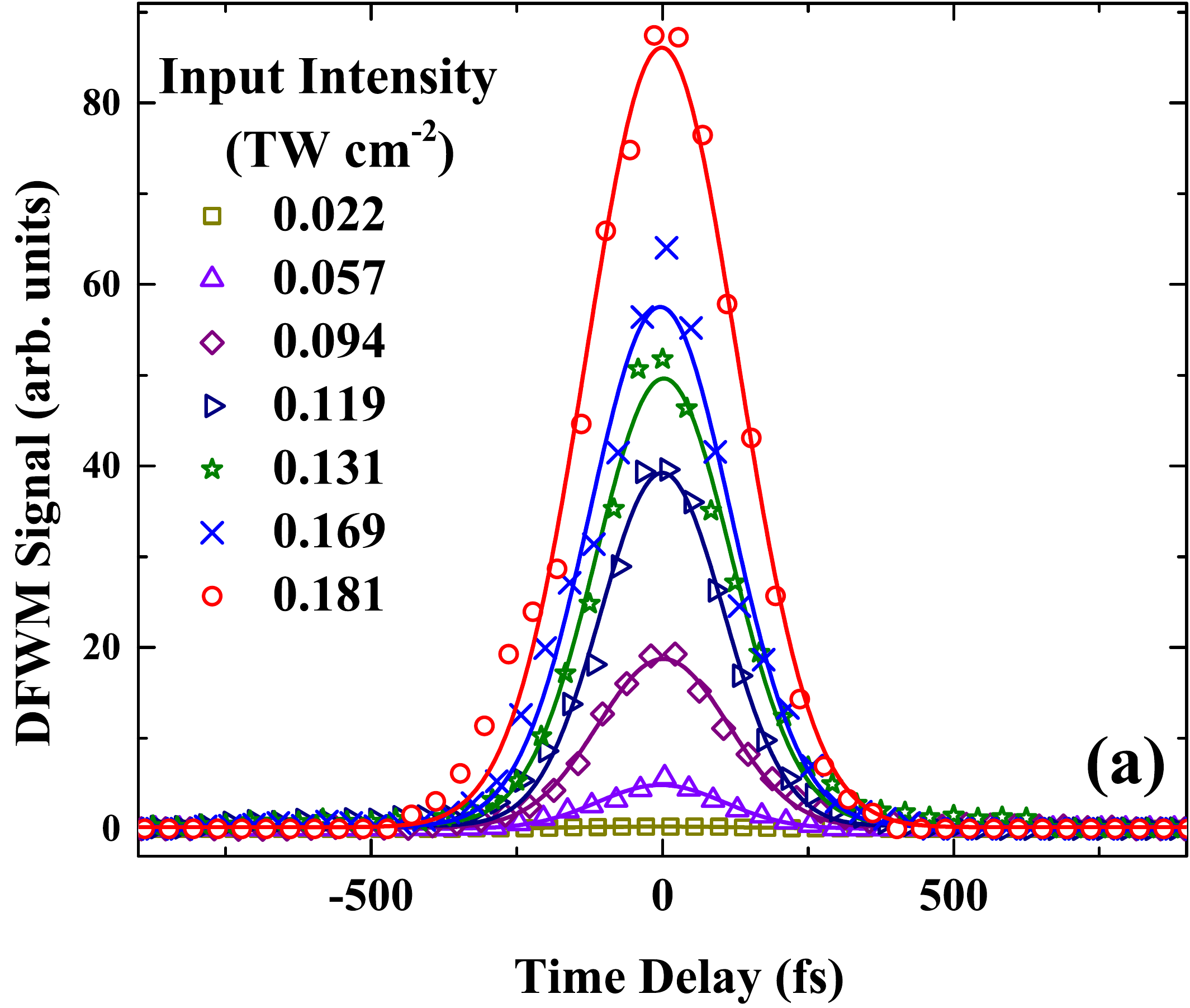}
  \includegraphics[height=0.25\textheight,keepaspectratio]{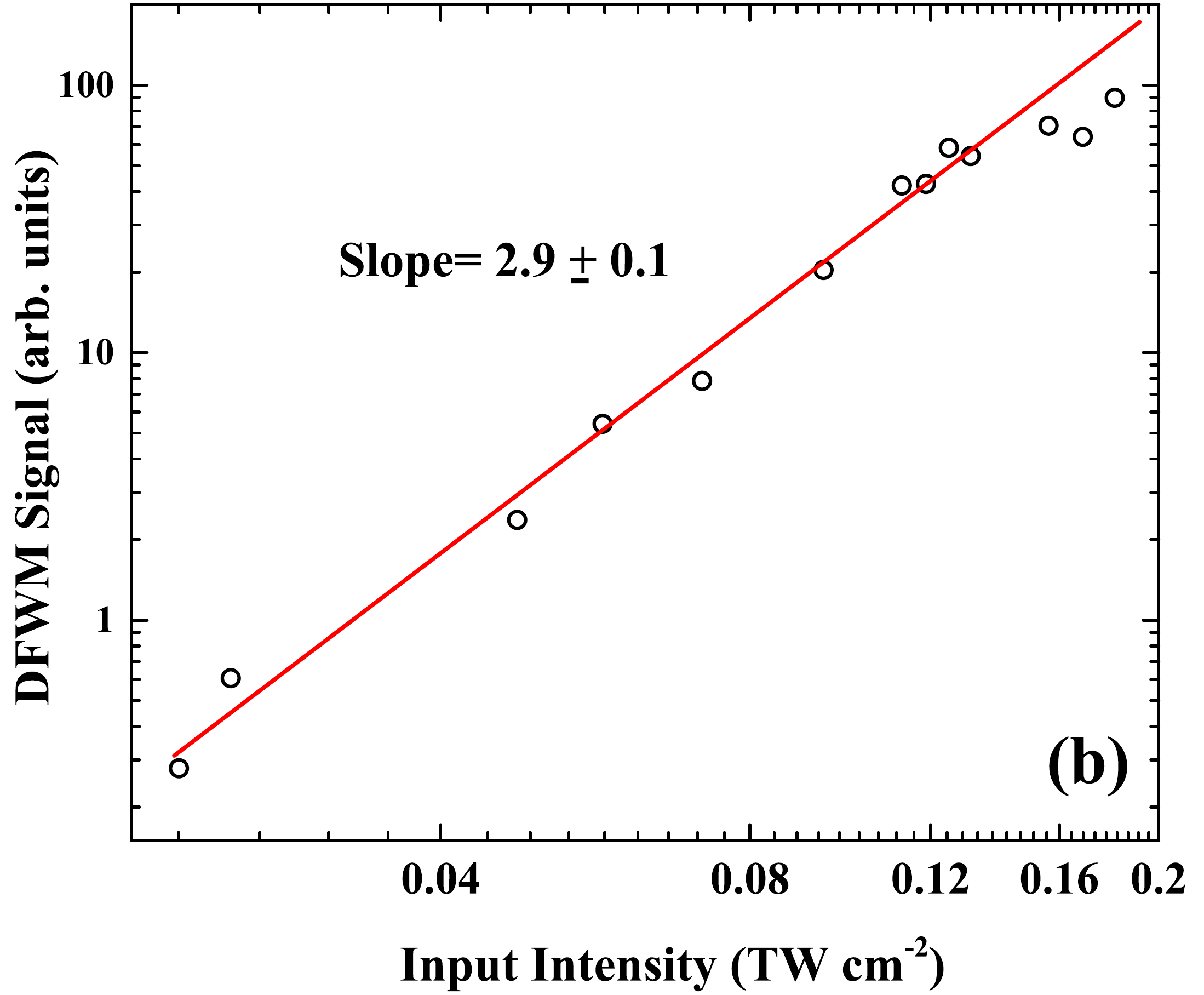}
\caption{\label{cds_dfwm_800nm} \footnotesize{(a) DFWM signal versus delay plots for different input intensity for 
colloidal solution of CdS QDs. The continuous lines are the Gaussian fitting.
(b) DFWM signal versus incident beam intensity plot.}}
\end{figure}

DFWM signal versus probe delay plots for 
colloidal solution of CdS QDs are shown in Fig. \ref{cds_dfwm_800nm} (a). 
The signals are fitted with Gaussian function (solid curve). 
The signal profiles are nearly symmetric about the
maximum ($\it i.e.$ zero time delay) illustrating that the response times
of the nonlinearities are shorter than the pulse duration (110 fs).  
This fast response enhances their potential for photonic switching applications. 
The intensity dependence of the DFWM signal amplitude is presented in Fig. \ref{cds_dfwm_800nm} (b). 
At relatively low input
intensities ($< 200$ GW/cm$^2$), the DFWM signal amplitude followes
a cubic (with a slope of 2.9$\pm$0.1) dependence.
It clearly demonstrates that the nonlinearity behaves in a Kerr-like fashion 
and the origin of DFWM does not have contribution from any
multiphoton absorption process, which leads to higher power dependence \cite{Sutherland1996}. 
It can be seen from the intensity dependence of the DFWM signal plot that 
the DFWM signal intensity goes down at input intensity around 180 GW/cm$^2$. 
This substantial reduction in the DFWM signal intensity is mainly due to 
the nonlinear absorption of all interacting beams. However, 
the DFWM signal does not show any higher power dependence, expected for multiphoton absorption, 
indicating the dominance of $\chi^{(3)}$ process over multiphoton photon absorption at this input intensity regime.
To confirm this, we have performed 
open-aperture Z-scan experiment, which is discussed in the next section.
The measurement of $\chi^{(3)}$ values are performed at zero time delay of all the beams.
We estimated the magnitude of $\chi^{(3)}_{1111}$ by maintaining the same polarization 
for all the three incident beams.
The third-order nonlinear optical susceptibility $\chi^{(3)}$ is estimated by comparing 
the measured DFWM signal of the sample with that of $CS_2$ as 
reference ($\chi^{(3)} = 5 \times 10^{-13}$ esu \cite{HBLiao1998, Minoshima1991}) measured with the same 
experimental conditions. The equation relating $\chi^{(3)}_{ref}$ and $\chi^{(3)}_{samp}$ 
is given by \cite{Sutherland1996} 
\begin{equation}
 \chi^{(3)}_{samp}=\Bigg(\frac{n_{samp}}{n_{ref}}\Bigg)^2\Bigg(\frac{I_{samp}}{I_{ref}}\Bigg)^{1/2}
 \Bigg(\frac{L_{ref}}{L_{samp}}\Bigg)\alpha L_{samp}
 \Bigg(\frac{e^{\frac{\alpha L_{samp}}{2}}}{1-e^{-\alpha L_{samp}}}\Bigg)\chi_{ref}^{(3)},
\label{chi_3_DFWM}
\end{equation}
where $I$ is the DFWM signal intensity, $\alpha$ is the linear absorption coefficient, 
$L$ is sample path length, and $n$ ($n_{samp}=1.329$ and $n_{CS_2}=1.606$ at $\lambda= 800$ nm) 
is the refractive-index. 
The effective refractive-index of the sample is essentially that of water solution.
The $\chi^{(3)}$ value for the  CdS QD sample comes out to be $(4.15\pm0.42) \times 10^{-13}$ esu 
for an input intensity of
47.5 GW/cm$^2$. 
Assuming no QD-QD interaction, the measured $\chi^{(3)}$ can be written as
\begin{equation}
 \chi^{(3)}=\Xi^{(3)} N,
\end{equation}
where $N$ is the QD concentration in the solution and $\Xi^{(3)}$ is the 
average nonlinearity per QD. The QD concentration for CdS QD sample is 3.2 M. 
The $\Xi^{(3)}$ value for the CdS QDs comes out to be 
$2.15 \times 10^{-30}$ esu per QD.
One of the main sources of error that arises in
experiments is through the intensity fluctuations of laser pulses. This
problem is tackled by taking the averaged data of 1000 pulses. The
second major source of error could be from the determination of solution concentration. 
Considering all the unforced random experimental errors, 
we estimate an overall error of 10 $\%$ in our calculations by repeating the experiments few times. \par
\begin{figure}[H]
  \centering
  \includegraphics[height=0.25\textheight,keepaspectratio]{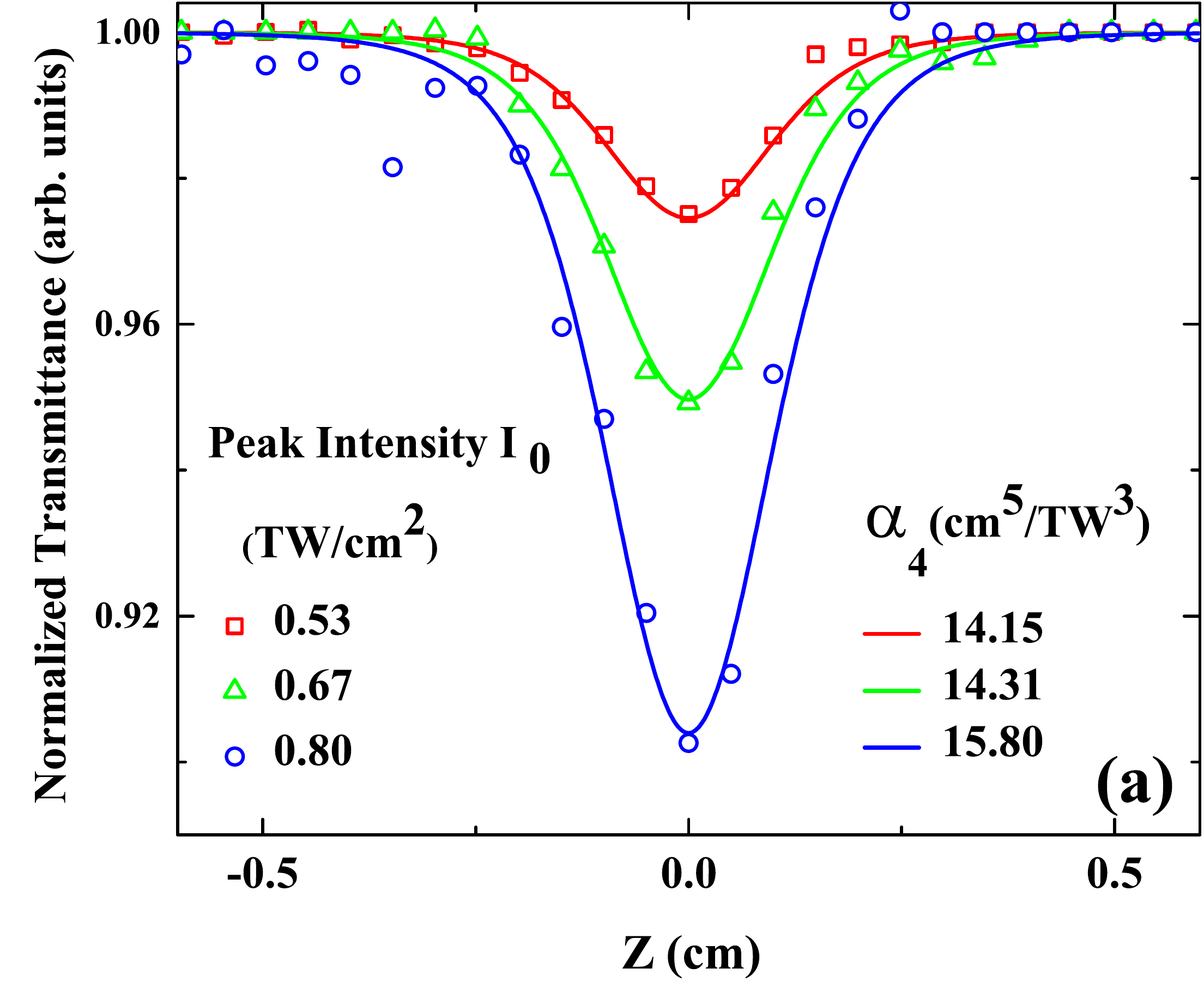}
  \includegraphics[height=0.25\textheight,keepaspectratio]{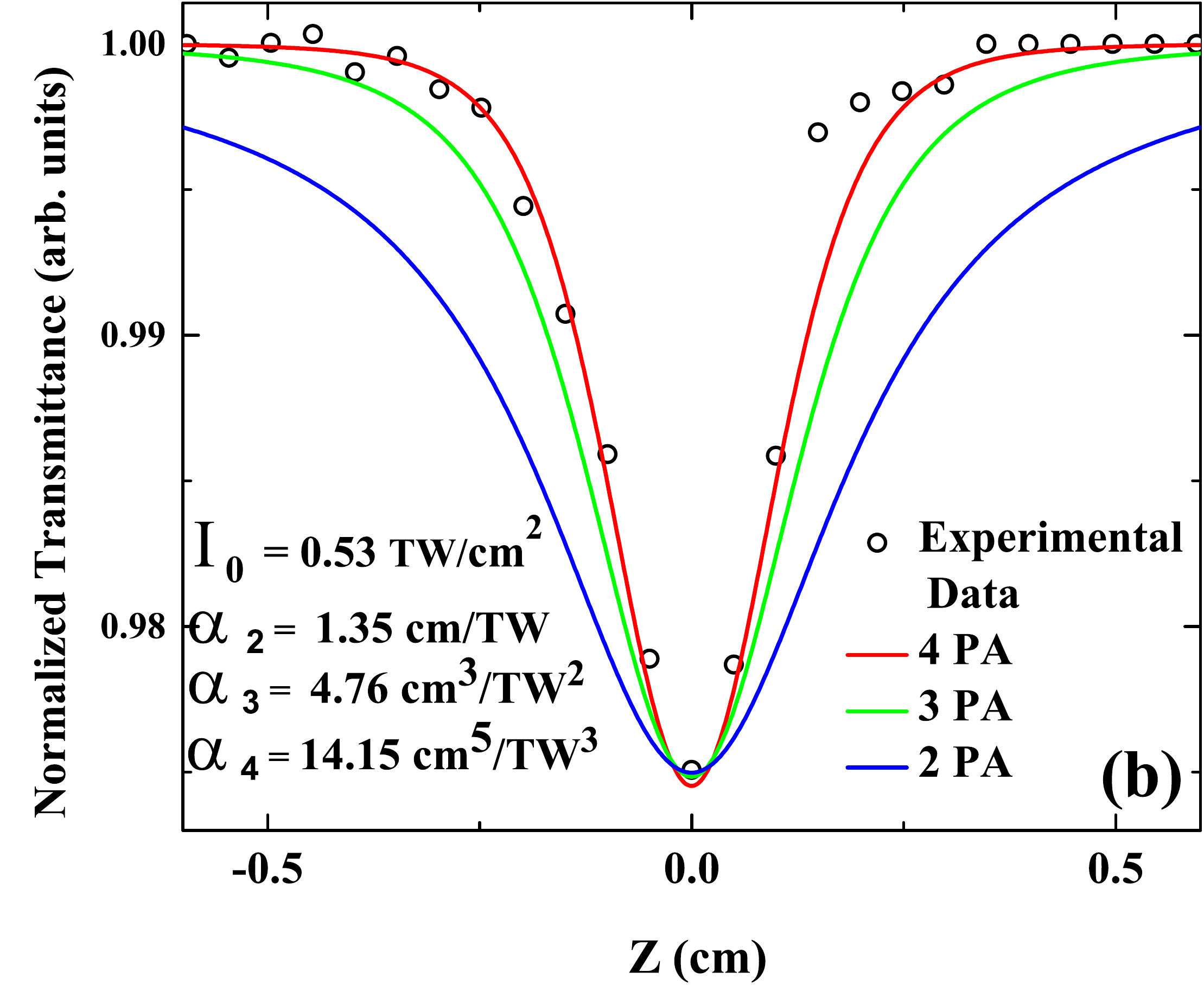}
  \caption{\label{oazs_800nm} \footnotesize{(a) Measured open-aperture Z-scan plots of 
  colloidal solution of CdS QDs for 800 nm wavelength, 110 fs laser pulses with different 
  input peak intensity (0.53 TW/cm$^2$, 
  0.67 TW/cm$^2$ and 0.80 TW/cm$^2$). The scattered points are experimental data points 
  and the continuous curves are the theoretical fitting corresponding to 4PA (n = 4). 
  (b) Theoretical fitting of the open-aperture Z-scan data 
  (corresponding to input peak intensity 0.53 TW/cm$^2$) considering TPA, 3PA, and 4PA.}}
\end{figure}
\begin{figure}[H]
  \centering
  \includegraphics[height=0.25\textheight,keepaspectratio]{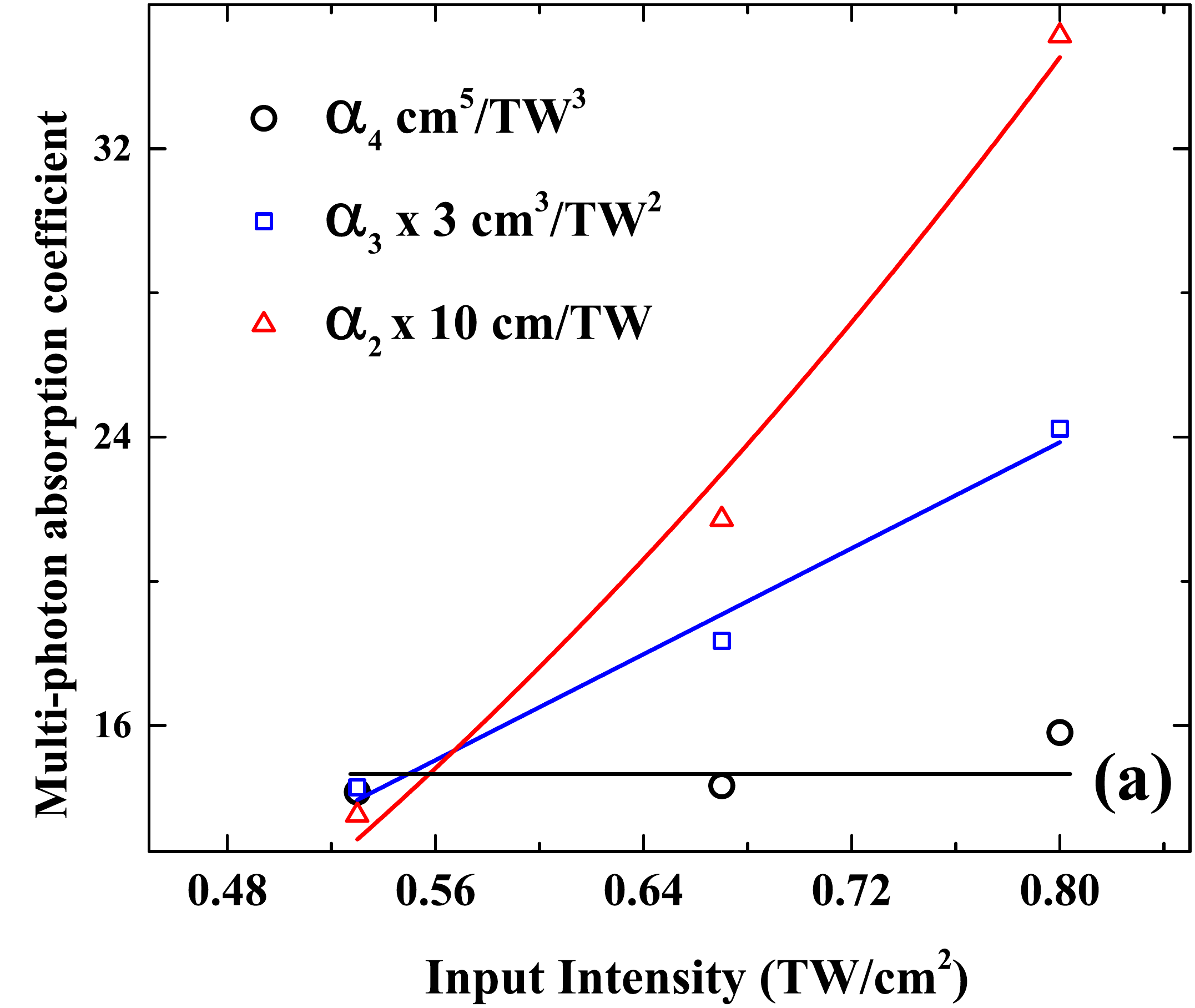}
  \includegraphics[height=0.25\textheight,keepaspectratio]{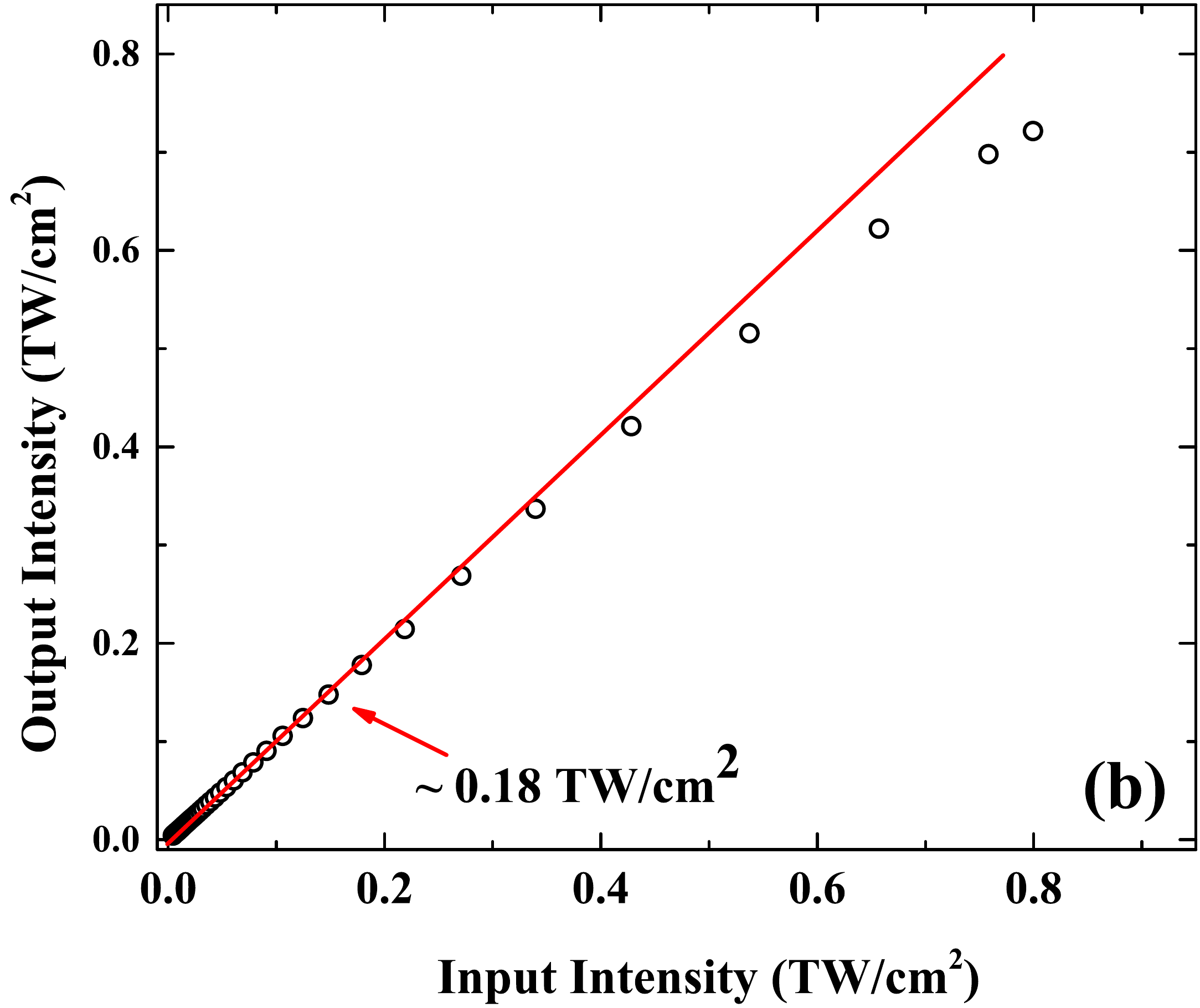}  
  \caption{\label{alpha234_OL_FSL}  \footnotesize{(a) Multi-photon absorption coefficient versus 
  incident beam intensity plots. 
  (b) Normalized transmittance plot for CdS QD sample. The red straight 
  line represents linear transmission.}}
\end{figure}
\begin{figure}[H]
  \centering
  \includegraphics[height=0.25\textheight,keepaspectratio]{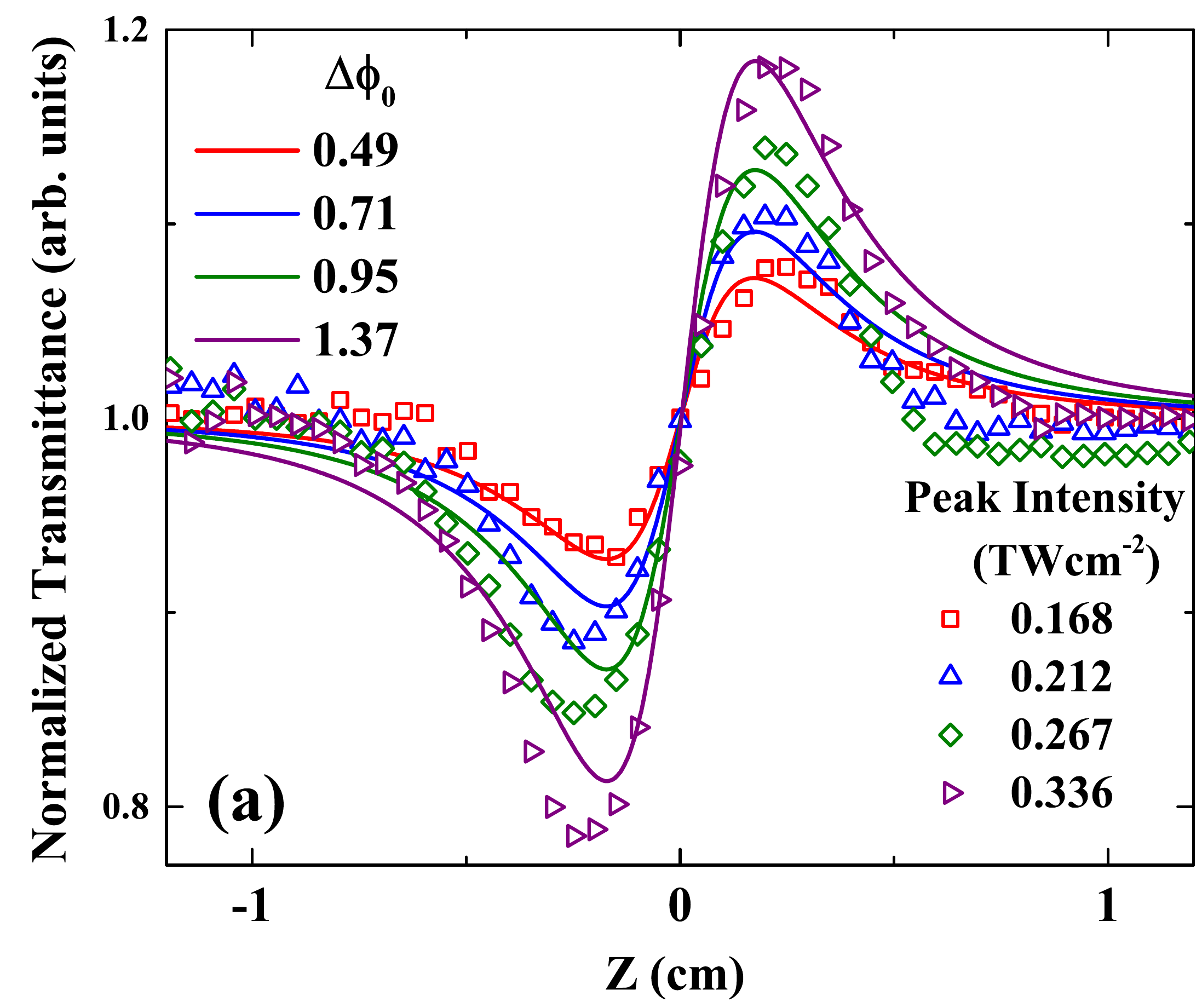}
  \includegraphics[height=0.25\textheight,keepaspectratio]{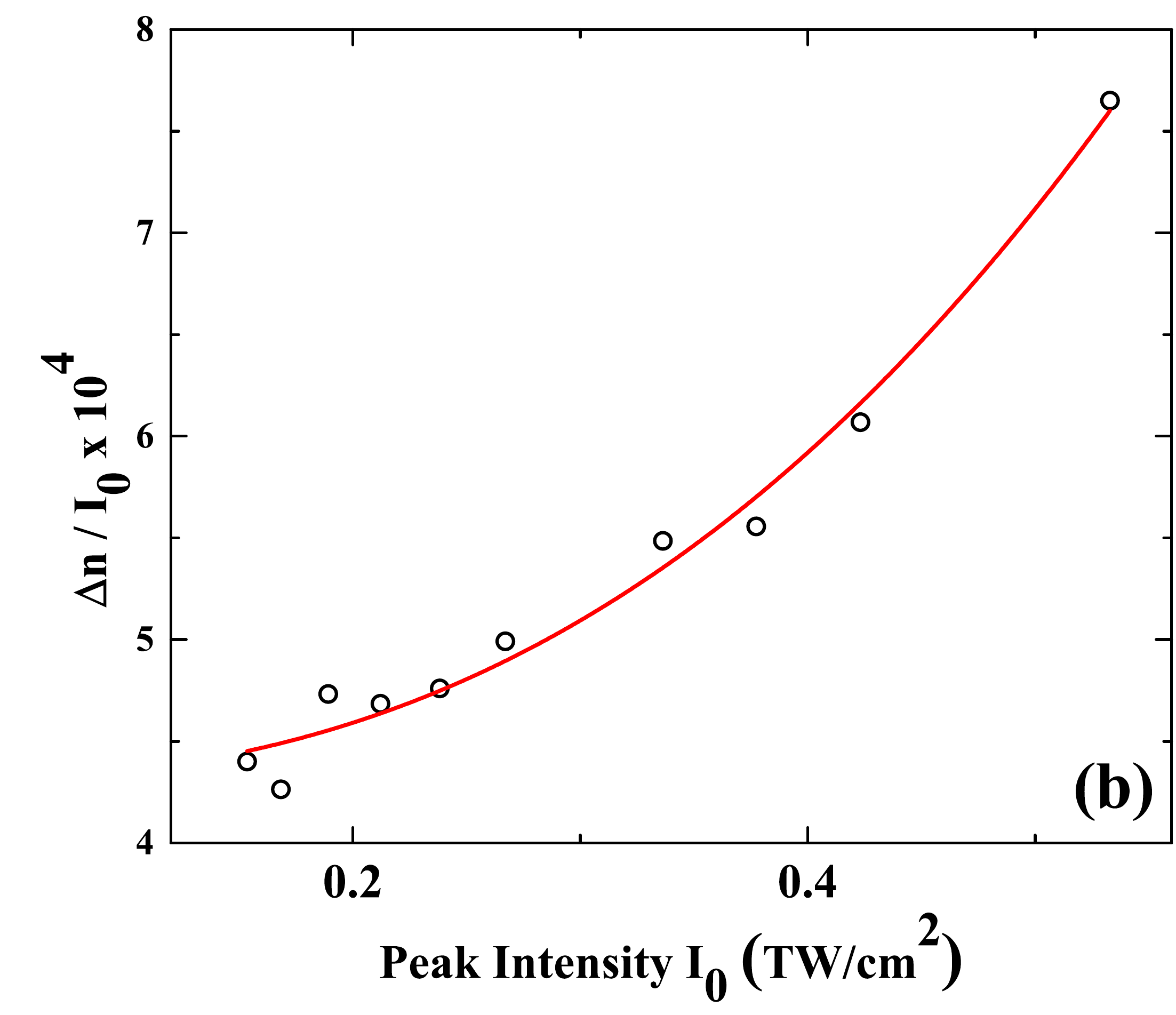}  
  \caption{\label{CAZS_ALL_deltan_vs_Intensity}  \footnotesize{ (a) Measured closed-aperture Z-scan plots with theoretical 
  fitting corresponding to different incident laser beam intensity for CdS QD sample. 
  The symbols represent experimental data points and the continuous curves are the theoretically 
  obtained plots.
  (b) $\Delta n/I_0$ versus $I_0$ plot for CdS QD sample. The continuous curve is the fitting following 
  Eq. \ref{gamma_sigma_r}.}}
\end{figure}

Towards performing Z-scan experiments, the incident Gaussian laser beam was passed through an aperture of 
diameter 3 mm and focused by a lens of focal length 12 cm. 
The beam waist ($\omega_0$) at the focal point (Z = 0) and the Rayleigh range ($Z_0 = k \omega_0^2/2$) 
were 23.3 $\mu$m and 2.13 cm, respectively. 
Whereas, the sample cell thickness was 1 mm.
Therefore, the sample was considered as 'thin' and the slowly varying envelope approximation (SVEA) 
was applied to obtain theoretical fitting of the experimental data points \cite{Sheik-Bahae1990}. 
Fig. \ref{oazs_800nm} (a) shows the measured open-aperture Z-scan plots of colloidal solution of CdS QDs 
for 800 nm wavelength, 110 fs laser pulses with three different input peak intensities (0.53 TW/cm$^2$, 
0.67 TW/cm$^2$, and 0.80 TW/cm$^2$). The scattered points are experimental data points 
and the continuous curves are the theoretical fitting corresponding to 4PA. 
All the theoretical simulations were performed following the analytic expression for open-aperture
Z-scan transmittance under first-order approximation given by Bing Gu {\it et al.} \cite{BingGuJOSA2010}. 
Fig. \ref{oazs_800nm} (b) shows the theoretical fitting of the open-aperture Z-scan data corresponding to 
input peak intensity 0.53 TW/cm$^2$ with n = 2, 3 , and 4.
The theoretical fitting obtained with n = 2 and 3 corresponding to 
two-photon (TPA) and three-photon absorption (3PA) 
do not exactly reproduce the experimental data. 
This is a clear indication that the 
the TPA and 3PA are not the dominant processes at 800 nm excitation. 
The curves are therefore  fitted with theoretically simulated result corresponding to four-photon absorption (4PA) 
process. 
The theoretical fitting with 4PA matches well with the experimental data. 
Fig. \ref{alpha234_OL_FSL} (a) shows multi-photon absorption coefficient versus 
incident beam intensity plots. 
It can be noted that the 
$\alpha_4$ value remains almost constant for the intensity range 0.53 TW/cm$^2$ to 0.80 TW/cm$^2$. 
Whereas, $\alpha_2$ and $\alpha_3$ increase quadratically 
and linearly with incident beam intensity, respectively. Therefore, it can be concluded that at 
this incident beam intensity range, four-photon absorption process is dominant.
Fig. \ref{alpha234_OL_FSL} (b) shows nonlinear transmittance plot for CdS QD sample. 
It shows that the nonlinear absorption starts at peak intensity around 0.18 TW/cm$^2$, 
which supports the results obtained in DFWM experiments.
Towards understanding the role of 4PA in nonlinear refraction, in case of excitation of 
CdS QDs with 800 nm femtosecond 
laser pulses, closed-aperture experiment was performed at different irradiances, 
ranging from 0.17 TW/cm$^2$ to 0.53 TW/cm$^2$.
Fig. \ref{CAZS_ALL_deltan_vs_Intensity} (a) shows the theoretical fitting 
of closed-aperture Z-scan plots corresponding 
to different incident beam intensity for CdS QD samples. 
All the theoretical fittings of closed-aperture
Z-scan transmittance results were performed following the analytic expression 
given by Bing Gu {\it et al.} \cite{BingGuJOSA2010}. 
The valley-peak configuration of the closed-aperture Z-scan curve 
indicates positive (self-focusing) nonlinearity due to the electronic Kerr-effect and 
excited state electrons reached by 4PA process. 
The corresponding phase equation can be given by \cite{Sheik-Bahae1990}
\begin{equation}
\frac{d\Delta \phi}{dz}= k\Delta n, 
\label{delta_n}
\end{equation}
where $\Delta n = \gamma I+\sigma_r N$ is the change in index of refraction. 
$\gamma$ is the nonlinear index corresponding to the bound electrons and $\sigma_r$ is the 
change in the refractive-index per unit photo-generated excited state electron density N. 
In the context of excited state electron generation due to 4PA, we can neglect excited state relaxation 
as these processes occur at longer time scale than the femtosecond laser 
pulses used for performing these experiments. 
Therefore, neglecting relaxation loss, the excited state electron generation rate due to 4PA can be given by 
\begin{equation}
\frac{d N}{dt}= \frac{\alpha_4 I^4}{4\hbar \omega}.
\label{carrier_generation_rate}
\end{equation}
Using Eqs. \ref{delta_n} and \ref{carrier_generation_rate}, we obtained the formula 
relating $\Delta n/I_{0}$ and $I_{0}$ for the presence of third-order nonlinearity and 
photo-generated excited state electrons by 4PA. 
The equation is given by 
\begin{equation}
 \Delta n/I_{0}=\gamma + C \sigma_r I_0^3,
 \label{gamma_sigma_r}
\end{equation}
where $C=0.23 (\alpha_4 \tau_0/4 \hbar \omega)$. 
Here $\tau_0$ is pulse width of the excitation laser beam.
In absence of nonlinear absorption, the difference between peak and valley ($\Delta T_{p-v}$) 
in closed-aperture Z-scan transmittance can be given by \cite{Said1992}
\begin{equation}
 \Delta T_{p-v}=p^{(3)} <\Delta \phi_0>,
\end{equation}
where $p^{(3)} = 0.406(1-S)^{0.25}$ and $\Delta \phi_0$ is the on-axis phase change at the focus. 
A closed and an open-aperture Z-scan are performed at same irradiance, and the closed-aperture data are divided 
by the open-aperture data. 
$\Delta T_{p-v}$ is obtained from the resultant curve. 
This value is then divided by $p^{(3)}k L_{eff} I_{0}/2^{1/2}$ to determine $\Delta n/I_0$. 
For determining $L_{eff}$, $\alpha$ is calculated using the formula $\alpha = \alpha_0 + \alpha_4 I_0^3$, 
where $\alpha_0$ is the linear absorption coefficient, and $\alpha_4$ is the 4PA coefficient which is 
obtained from the open-aperture Z-scan experiment results. 
The experiments are performed at different irradiances, and $\Delta n/I_{0}$ is plotted 
as function of $I_0$. 
$\Delta n/I_{0}$ versus $I_{0}$ plot is shown in Fig. \ref{CAZS_ALL_deltan_vs_Intensity} (b). 
In absence of any higher-order nonlinearity, this plot is expected to be a horizontal 
line with vertical intercept $\gamma$. 
From the theoretical fitting (red continuous curve) using Eq. \ref{gamma_sigma_r}, 
$\gamma$ and $\sigma_r$ are calculated and the values are $(4.45 \pm 0.1) \times 10^{-4}$ $cm^2/TW$ and 
$(6.0 \pm 0.3) \times 10^{-21}$ cm$^3$, respectively. 
Therefore, the closed-aperture Z-scan results 
further establish the 4PA processes.\par
\begin{figure}[H]
  \centering
  \includegraphics[height=0.25\textheight,keepaspectratio]{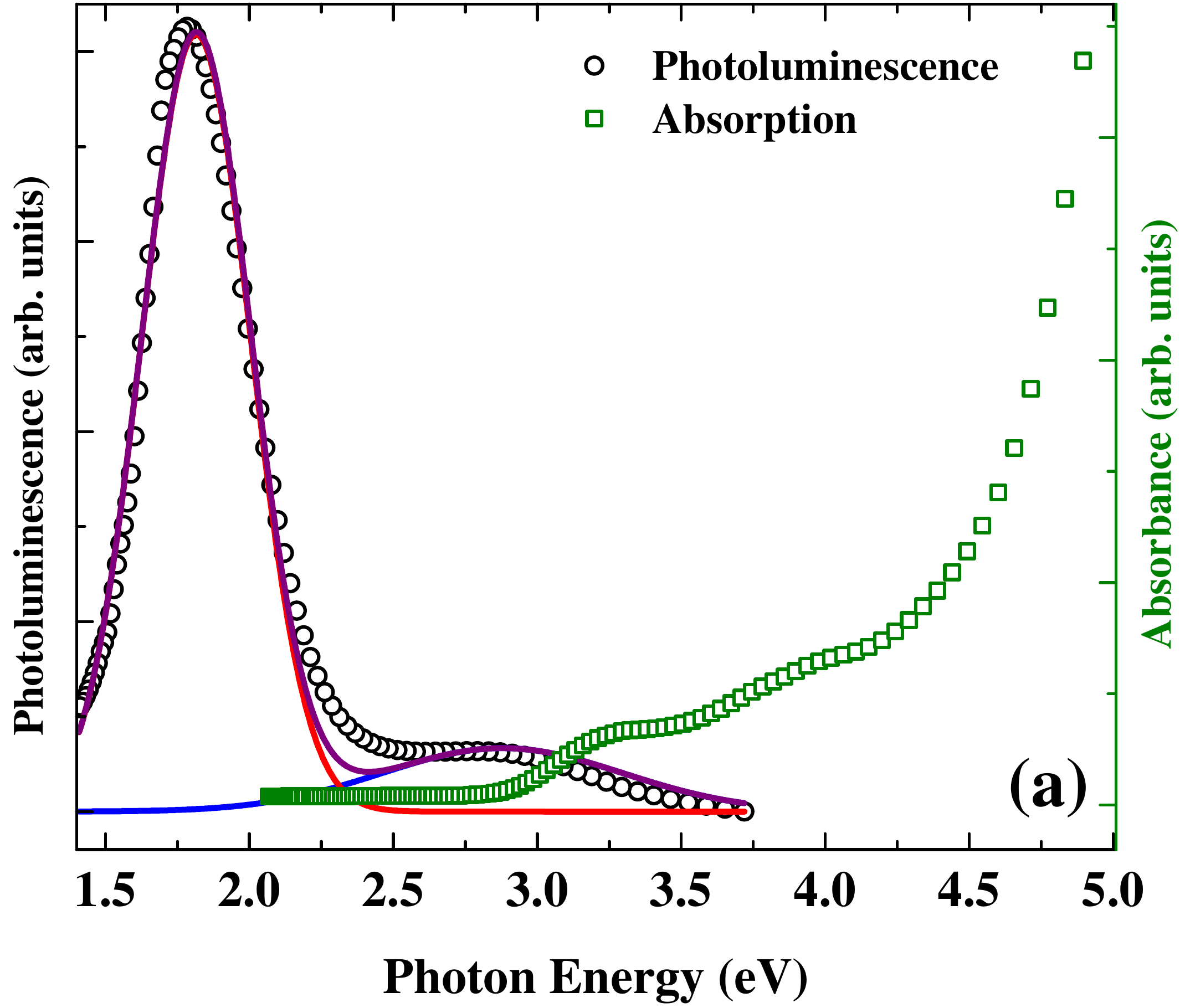}
  \includegraphics[height=0.25\textheight,keepaspectratio]{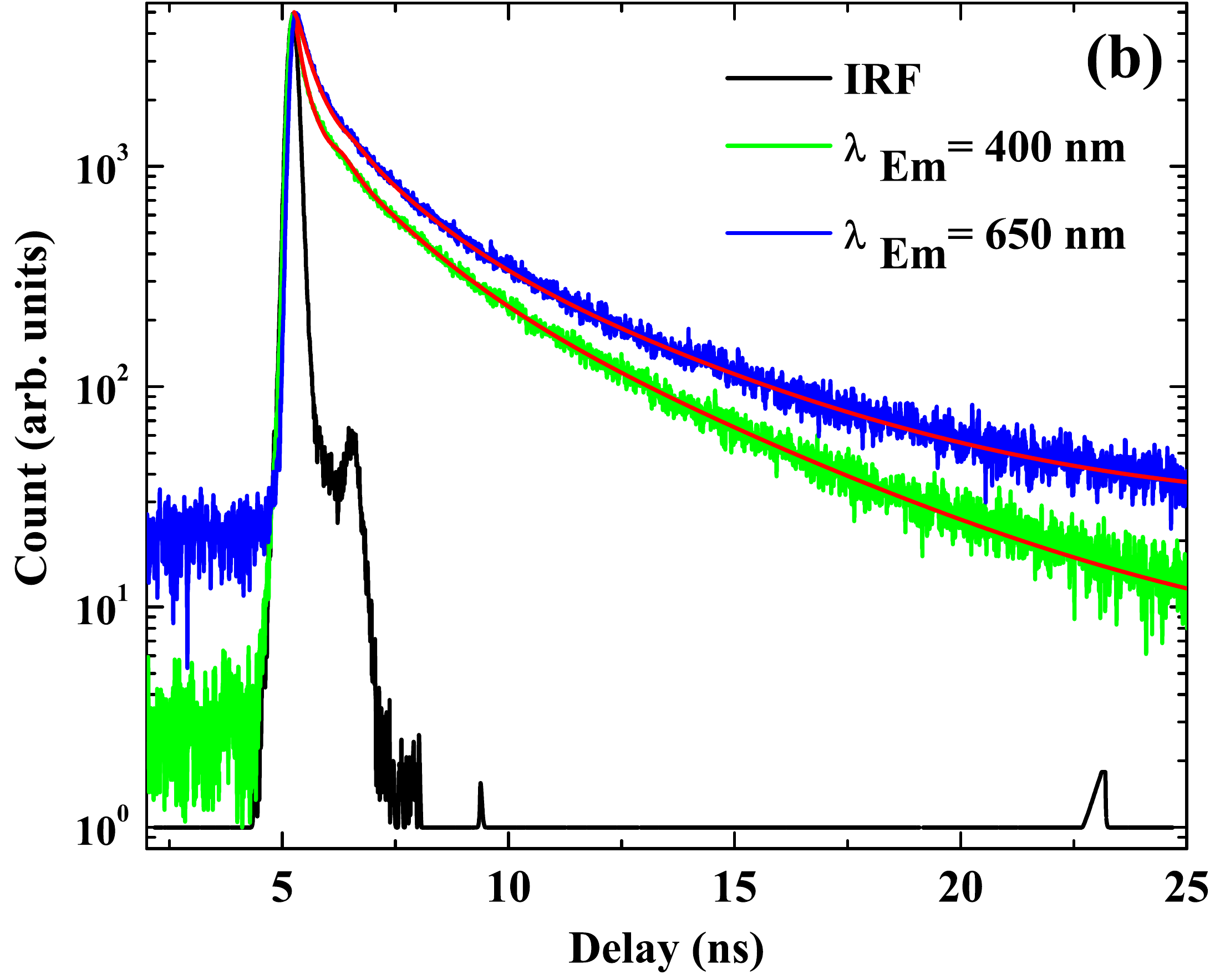}
  \caption{\label{Abs_PL_TCSPC} \footnotesize{ (a) Absorption and photoluminescence spectra
  of the colloidal solution of CdS QDs. The PL spectrum is fitted with two Gaussians.
  The black circles are experimental data points and continuous violet line
  is the cumulative fit. 
  (b) Fluorescence decay curves corresponding to monitoring wavelengths 400 nm and 650 nm 
  for the colloidal solution of CdS QDs.}}
\end{figure}
\begin{figure}[H]
 \centering
  \includegraphics[height=0.28\textheight,keepaspectratio]{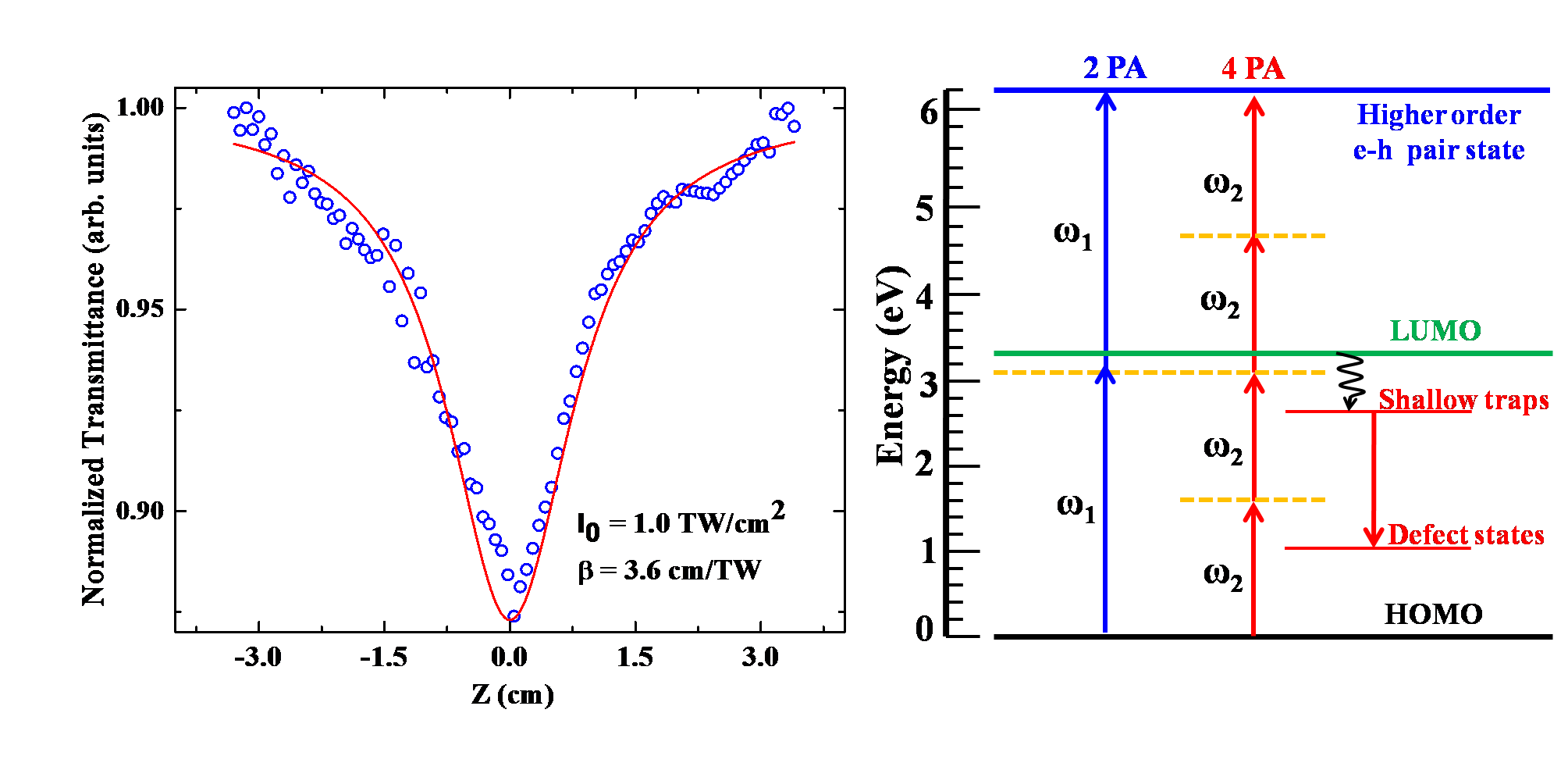}
  \caption{\label{OAZSCAN_400nm_electronic_states_schematic}  \footnotesize{ Left panel) Open-aperture Z-scan with 
  400 nm femtosecond laser pulses. Right panel) Quasi-molecular orbital energy levels (horizontal solid lines), 
  localized defect states and electronic transitions (perpendicular lines with arrows) 
  corresponding to TPA and 4PA in CdS QDs. The dashed lines represent intermediate virtual electronic states.}}
\end{figure}

In this section, the results of the linear studies including UV-visible absorption, 
room temperature photoluminescence, and TCSPC are reported towards establishing 
the energetic positions of the electronic states and their decay times.
Fig. \ref{Abs_PL_TCSPC} (a) shows absorption and photoluminescence spectra 
of colloidal solution of CdS QDs. 
The peak positions of the absorption band-edge for these semiconductor QD sample appears at 
around 380 nm wavelength ($\sim$ 3.26 eV). 
Whereas, the band gap of bulk CdS is 2.42 eV. 
This large blue shift of absorption band-edge is due to quantum confinement effect in 
QDs having diameter less than 5.8 nm (Bohr radius of bulk CdS). 
The broadness of the absorption band-edge suggests broad particle size 
distribution and confirmed by HRTEM images \cite{Soumyendu2012}. 
The average diameter of the 
QDs is 4.2 nm.
The photoluminescence spectrum of these CdS QDs manifests 
two broad bands corresponding to Stokes shifted band-edge emission 
and defect state emission. 
The band-edge photoluminescence band ranges from 350 nm ($\sim$ 3.5 eV) to 
500 nm ($\sim$ 2.5 eV). 
Whereas, the defect state emission band energy ranges from 2.5 eV to 
1.6 eV with peak at around 1.9 eV. 
Time correlated single photon counting (TCSPC) was performed 
to determine the decay time of the band-edge and defect-state transitions in these QD sample.
The fluorescence decay plots of the colloidal solutions of CdS QDs are shown 
in Fig. \ref{Abs_PL_TCSPC} (b).  
A picosecond laser of wavelength 375 nm is used as excitation source. 
The PL emission is monitored at wavelengths  400 nm and 650 nm which correspond 
to the band-to-band and defect state transitions respectively. 
The FWHM of the instrument respose function (IRF) is 254 ps.
The curves can be fitted with three exponential decay functions. 
The fluorescence decay times corresponding to 400 nm emission wavelength are: $\tau_1 \sim 0.2$ ns, 
$\tau_1 \sim 1.5$ ns, and $\tau_1 \sim 4.5$ ns. The decay times corresponding to 
650 nm emission wavelength are: $\tau_1 \sim 0.3$ ns, 
$\tau_1 \sim 1.4$ ns, and $\tau_1 \sim 4.5$ ns.
Whereas, the decay time obtained in DFWM with 800 nm femtosecond 
laser pulses is of the order of 110 fs. 
These results confirm that electrons do not get excited to band edge or defect states 
for 800 nm fs laser pulse excitation. 
Left panel of Fig. \ref{OAZSCAN_400nm_electronic_states_schematic} shows 
open-aperture Z-scan with 400 nm femtosecond laser 
pulses with peak intensity 1.0 TW/cm$^2$. 
The theoretically simulated result corresponding to two-photon absorption (TPA) 
adequately reproduces the experimental data. 
The TPA coefficient value comes out to be 3.6 cm/TW. \par
The schematic description of two-photon and four-photon transition processes and all the electronic states 
probed by linear and nonlinear optical techniques are shown in 
the right panel of Fig. \ref{OAZSCAN_400nm_electronic_states_schematic}.
The energy corresponding to 3PA for 800 nm excitation wavelength is 4.65 eV. 
Whereas, Dhayal {\it et al.} \cite{Dhayal2014} have reported that there are real excited states at this 
energy level for CdS QDs. Therefore, it can be concluded that the electrons absorb three-photons initially and 
reach these real excited states. 
Thereafter the electrons absorb one more photon to reach the terminal state. 
Therefore, this multiphoton absorption process can be called as 3PA assisted 4PA.
\section{Conclusion}
The ultrafast nonlinear optical properties
including the time response of CdS QD sample 
using degenerate four-wave mixing technique at a wavelength of 800 nm with 110 fs pulses were 
thoroughly investigated. 
The nonlinear experiments were performed for the intensity 
regime 0.02 TW/cm$^2$ to 0.80 TW/cm$^2$. The CdS QD sample shows Kerr-type nonlinearity for intensity 
below 0.18 TW/cm$^2$. However, the intensity dependent open-aperture and 
closed-aperture Z-scan studies with 800 nm femtosecond 
laser pulses indicate 4PA above this input intensity. 
The closed-aperture Z-scan also manifests positive nonlinearity (self-focusing) 
for the CdS QDs.
Open-aperture Z-scan with 400 nm femtosecond laser pulses shows two-photon absorption (TPA). \par
Band gap energy and the defect state energy of the CdS QDs were estimated from the UV-visible 
absorption and PL spectrum. 
Whereas, information about the energy positions of the higher-order 
electronic states is obtained from the multiphoton absorption processes. 
\section*{Acknowledgments}
We thank Prof. A. S. Pente,  
BARC, Mumbai, for providing gamma-irradiation facility for synthesis of CdS QDs. 
We also thank CRNTS and Central Surface Analytical Facility, IIT Bombay for facilitating HR-TEM characterizations.
\section*{References}
\bibliographystyle{apsrev}
\bibliography{REFERENCES_pin3}
\end{document}